\documentclass[12pt]{iopart}
\usepackage{array}
\usepackage{epsfig}
\usepackage{amssymb}
\usepackage{graphicx}
\usepackage{srctex}
\usepackage{cite}
\usepackage[usenames]{color}
\usepackage{dcolumn}
\usepackage{bm}

\usepackage{iopams}  
\begin{document}

\title[ ]{Equivalence of the modified Villain formulation and the dual Hamiltonian method in the duality of the XY-plaquette model }

\author{Morishige Yoneda}

\address{Aichi University of Technology, 50-2,
Umanori Nishihasama-cho , Aichi 443-0047, Japan.}
\ead{yoneda-morishige@aut.ac.jp}
\vspace{10pt}
\begin{indented}
\item[] \today 
\end{indented}

\begin{abstract}
Regarding the duality of the XY-plaquette model in the $2+1d$ system, We compared and discussed the equivalence and difference between the "modified Villain formulation (MVF)", recently introduced by Gorantla et al. to study duality in exotic field theories such as fracton theory, and the "standard Villain formulation (SVF) and dual Hamiltonian method (DHM)" we introduced in our previous paper to study duality between Josephson junctions (JJ) and quantum phase slips (QPS) in the $1+1d$ system.
\end{abstract}

\section{Introduction}
 Duality has a very broad meaning and plays a very important role in statistical mechanics and field theory.
 One typical example is to relate the temperature of the original system (the coupling constant) to the inverse temperature of the dual system (the inverse of the coupling constant). Another typical example is electromagnetic duality due to the exchange of electric and magnetic fields by gauge theory. Specifically, it is the Kramers-Wanier duality in the two-dimensional Ising model \cite{ref1}, Montonen-Olive duality in the two-dimensional Ising model \cite{ref2}, Montonen-Olive duality in field theory \cite{ref2}, and S-duality \cite{ref3} and T-duality \cite{ref4} in superstring theory are known. As dual transformations in the XY model and U(1) lattice gauge theory \cite{ref5,ref6}, the Villain approximation and the Villain model dual transformation \cite{ref7,ref8,ref9,ref10} are well known. Recent topics related to the Villain model, the Modified Villain formulation (MVF) \cite{ref11,ref12,ref13,ref14,ref15} has received renewed attention for studying duality, such as the U(1) gauge theory by Sulejmanpasic et al. \cite{ref11,ref12} and the fracton theory \cite{ref13,ref14,ref15,ref16,ref17,ref18,ref19,ref20} in exotic field theories by Gorantla et al. \cite{ref13,ref14,ref15}. Note that while this formulation is in principle valid for $1+1d$ systems, it also gives an exact self dual transformation in the XY-plaquette model \cite{ref21}, which is a $1+2d$ system. In this paper, we call a system that appears to be accompanied by dimensionality reduction a "pseudo-$1+1d$ system", as in the example of the XY-plaquette model, where the system has more than $1+1d$ dimensions but exhibits the same dual transformation properties as the $1+1d$ system. The purpose of this paper is to compare and contrast, and to discuss the equivalence and differences, the " MVF " introduced by Gorantla et al. and the "standard Villain formulation (SVF)" and "dual Hamiltonian method (DHM) \cite{ref22,ref23}" introduced to study the duality in our previous paper between Josephson junctions(JJ) and quantum phase slips junction (QPSJ) \cite{ref24,ref25,ref26,ref27} in $1+1d$ systems. This paper is organized as follows. In Section 2, we treat the dual transformation of the XY-plaquette model as an example for the case of a pseudo $1+1d$ system, following the MVF by Gorantla et al. In Section 3, we show that the dual transformation using the SVF in the XY-plaquette model has exactly the same result as the MVF in Section 2. In Section 4, we show that it is possible to construct an exact self dual system for the Villain approximation version of the XY-plaquette model without using Poisson's summation formula \cite{ref9,ref10} by implementing the algorithm of the DHM. In Section 5, provides a summary and discussion. In Appendix A, we review the Dual transformation using the MVF for the $1+1d$ XY-model. In Appendix B, we review the momentum symmetry current and winding symmetry current of the MVF in the $1+1d$ XY-model. In Appendix C, we describe an example in which the DH method is introduced to construct a self duality \cite{ref22} between the JJ and the QPSJ. In Appendix D, we show that the dual transformation results from the SVF in the $1+1d$ XY-model are self dual as well as those from the two-component MVF.
\section{Dual transformations by MVF in pseudo-$1+1d$ system}
In this section, we treat the dual transformation of the XY-plaquette model \cite{ref14,ref15,ref21} as an example for the case of a pseudo-$1+1d$ system, following the MVF by Gorantla \cite{ref14,ref15} et al.
\subsection{Dual transformations in XY-plaquette model by MVF}
In this subsection, referring to the MVF in the XY-model of the $1+1d$ system presented in Appendix A, we show that the XY-plaquette model in the $1+2d$ system is a pseudo-$1+1d$ system where it undergoes the same dual transformation as the XY model of the $1+1d$ system. As a starting point, the partition function for the lattice version of the $1+2d$ XY plaquette model \cite{ref14,ref15,ref21} is as follows:
\begin{eqnarray}\label{eq1} 
 \scalebox{0.90}{$\displaystyle
Z\left( {\beta }_0,{\beta }_{xy} \right)=\int{\mathcal{D}\theta}\exp\sum\limits_{x,y,\tau }\Biggl\{{ -{{\beta }'_0}\left( 1-\cos {{\nabla }_{\tau }}\theta  \right)-{{\beta }'_{xy}}\left( 1-\cos {{\nabla }_{xy}}\theta  \right)}\Biggr\}
 $},\nonumber \\
 \scalebox{0.8}{$\displaystyle\hspace{80px}
 \int{\mathcal{D}\theta}\equiv {\prod\limits_{\tau =1}^{N_\tau}\prod\limits_{x=1}^{N_x}\prod\limits_{y=1}^{N_y}{\int\limits_{-\pi }^{\pi }\frac{\displaystyle d{\theta}\left( x, y,\tau \right)}{\displaystyle 2\pi }}}
 $},
\end{eqnarray}
where its difference operator is defined by ${\nabla }_{xy}\equiv {\nabla }_x{\nabla }_y$,  is the same as defined in Eq. (\ref{eqA1}) in Appendix A. ${\beta }_0$ and ${\beta }_{xy}$ are the energies of the imaginary time $\tau$ and the two-dimensional $xy$-space component, respectively, and ${\beta }'_0\equiv{\beta }_0\Delta \tau/{\hbar }\;$ and ${\beta }'_{xy}\equiv{\beta }_{xy}\Delta \tau /{\hbar }\;$ are the respective dimensionless energies. Similar to Eq. (\ref{eqA6}), the Villain approximation \cite{ref7,ref9,ref22,ref23} for Eq. (\ref{eq1}) is as follows: 
\begin{eqnarray}\label{eq2} 
\scalebox{0.80}{$\displaystyle\hspace{-60px}
Z_v\left( {\beta }_0,{\beta }_{xy} \right)\equiv R_{QV}\int\mathcal{D}\theta\sum\limits_{\left\{ n_{\tau },n_{xy} \right\}}{\exp }\sum\limits_{x,y,\tau }{\left\{ -\frac{{\left( {\beta }'_0 \right)}_v}{2}{\left({\nabla }_{\tau }\theta -2\pi n_{\tau } \right)}^2-\frac{{\left( {\beta }'_{xy} \right)}_v}{2}{\left( {\nabla }_{xy}\theta -2\pi n_{xy} \right)}^2 \right\}}
$},
\end{eqnarray}
Following the discussion in Eq. (\ref{eqA8})$-$(\ref{eqA17}) in Appendix A, adding an extra additional term +$i\tilde{\theta }\omega$ due to the dual phase field $\tilde{\theta }$ to Eq. (\ref{eq2}) and then applying the two-component MVF \cite{ref14}, we obtain the dual transformation as follows: 
\begin{eqnarray}\label{eq3}
 \scalebox{0.80}{$\displaystyle\hspace{-90px}
Z_{MV}\!\left({\beta }_0,{\beta }_{xy} \right)\!\equiv\!R_{QV}\!\!\int\!\mathcal{D}\theta\int\!\mathcal{D}\tilde{\theta }\sum\limits_{\left\{ n_i \right\}}\exp\!\left( -1 \right)\!\sum\limits_{x,\tau }{\left\{ \frac{\left({\beta }'_0 \right)_v}{2}{\left({\nabla }_{\tau }\theta\! -\!2\pi n_{\tau } \right)}^2\!+\!\frac{\left( {\beta }'_{xy} \right)_v}{2}{{\left( {\nabla }_{xy}\theta \!-\!2\pi n_{xy} \right)}^2}\!+\!i\tilde{\theta }{\omega }_{\tau,xy} \right\}}
$}\nonumber\\
\scalebox{0.80}{$\displaystyle\hspace{-80px}
=C_{QV}\int\mathcal{D}\theta\int\mathcal{D}\tilde{\theta }\sum\limits_{\tilde{n}_x,\tilde{n}_y}\!\exp\!\left( -1 \right)\!\sum\limits_{x,\tau }{\left\{ \frac{( \tilde{\beta }'_0 )_v}{2}( {\nabla }_{\tau }\tilde{\theta }\!-\!2\pi \tilde{n}_{\tau } )^2\!+\!\frac{(\tilde{\beta }'_{xy} )_v}{2}({\nabla }_{xy}\tilde\theta\!-\!2\pi\tilde{n}_{xy} )^2\!-\!i\theta\tilde{\omega }_{\tau,xy} \right\}}
 $},
\end{eqnarray}
where ${\omega}_{\tau,xy}$ and $\tilde{\omega }_{\tau,xy}$ are defined as curl of $n_i^{xy}$ and $\tilde{n}_i^{xy}$ respectively, as follows:
\begin{eqnarray}\label{eq4}
 \scalebox{1.0}{$\displaystyle\hspace{100px}
\omega_{\tau,xy}\equiv \varepsilon _{ij}^{xy}\nabla _i^{xy}n_j^{xy}
 $},
 \end{eqnarray}
\begin{eqnarray}\label{eq5}
 \scalebox{1.0}{$\displaystyle\hspace{100px}
\;{{\tilde{\omega }}_{\tau,xy}}\equiv \varepsilon _{ij}^{xy}\nabla _i^{xy}\tilde{n}_j^{xy}
 $},
 \end{eqnarray}
where $\nabla _0^{xy}$, $\nabla _1^{xy}$, $n_0^{xy}$, $n_1^{xy}$, $\tilde{n}_0^{xy}$, and $\tilde{n}_1^{xy}$ are defined by $\nabla _0^{xy}\equiv\nabla_{\tau }$, $\nabla _1^{xy}\equiv\nabla_{xy}$, $n_0^{xy}\equiv {{n}_{\tau }}$, $n_1^{xy}\equiv n_{xy}$, $\tilde{n}_0^{xy}\equiv\tilde{n}_{\tau }$ and $\tilde{n}_1^{xy}\equiv\tilde{n}_{xy}$ respectively, and, ${\varepsilon}_{01}^{xy}\equiv {\varepsilon }_{\tau,xy }=1$ and $\varepsilon _{10}^{xy}\equiv\varepsilon_{xy,\tau }=-1$ are pseudo-two-dimensional Levi-Civita symbols. The Villain's inverse dimensionless energies $( \tilde{\beta' }_0 )_v$ and $( \tilde{\beta'}_{xy} )_v$ are the ''cross duals'' to $\left( {\beta }'_{xy} \right)_v$ and   respectively, as in Eq. (\ref{eqA17}), and are as follows: 
\begin{eqnarray}\label{eq6}
 \scalebox{1.0}{$\displaystyle\hspace{100px}
{{( \tilde{\beta }'_0 )}_v}\equiv \frac{1}{4{{\pi }^2}{\left(\beta'_{xy}\right)}_v}
 $},
 \end{eqnarray}
\begin{eqnarray}\label{eq7}
 \scalebox{1.0}{$\displaystyle\hspace{100px}
( \tilde{\beta'}_{xy} )_v\equiv \frac{1}{4{{\pi }^2}{\left( \beta'_0 \right)}_v}
 $},
 \end{eqnarray}
\subsection{Noether and winding currents with momentum and winding symmetry in the XY-plaquette model}
In this subsection, we consider the Noether current for two symmetries \cite{ref18,ref19,ref20}, $U(1)$ momentum global symmetry and $U(1)$ winding global symmetry, according to Appendix B. Now, following references \cite{ref18,ref19,ref20}, we extract from the partition function on the left hand side of Eq. (\ref{eq3}) a dimensionless Lagrangian described in a subspace instead of the entire $1+2d$ space, as follows:
\begin{eqnarray}\label{eq8}
\scalebox{0.90}{$\displaystyle\hspace{40px}
L=-\frac{\left( {\beta' }_i^{xy} \right)_v}{2}\sum\limits_{plaquette}\left( \nabla _i^{xy}\theta -2\pi n_i^{xy} \right)^2+i\sum\limits_{cube}\varepsilon _{ij}^{xy}n_i^{xy}\nabla _j^{xy}\tilde{\theta }
$} \end{eqnarray}
where $\beta _0^{xy}$ and $\beta _{1}^{xy}$ are defined by $\beta _0^{xy}\equiv {\beta }_0$ and $\beta _1^{xy}\equiv \beta_{xy}$, respectively, and $\sum\limits_{link}$, $\sum\limits_{plaquette}$ and $\sum\limits_{cube}$ are the sum of link, plaquette and cube in subspace of $1+2d$ space, respectively. In the Lagrangian of Eq. (\ref{eq8}), as in Eq. (\ref{eqB2}), from the noether current due to the shift $\theta \to \theta +c^m$ of the real constant cm of the phase field $\theta$, called the $U(1)$ momentum global symmetry, the momentum current $J_i^{xy}\left( i=0,1 \right)$ \cite{ref18,ref19,ref20} is as follows: 
\begin{eqnarray}\label{eq9}
\scalebox{0.90}{$\displaystyle\hspace{80px}
J_i^{xy}\left( \mathbf{x},\tau  \right)=-i\left( {\beta' }_i^{xy} \right)_v\left( \nabla _i^{xy}\theta -2\pi n_i^{xy} \right)\left( \mathbf{x},\tau  \right)
$},
 \end{eqnarray}
where $\left( \mathbf{x},\tau \right)=\left( x,y,\tau  \right)$, and $J_0^{xy}$ and $J_1^{xy}$ are defined by $J_0^{xy}=J_{\tau }$ and $J_1^{xy}=J_{xy}$, respectively. Note that on the right-hand side of Eq. (\ref{eq9}), no contraction is taken for the index $i$. 
Next, we will discuss the according to Eq. (\ref{eqB4}), in the Lagrangian of Eq. (\ref{eq8}), the winding current $J_i^{w}$ due to $U(1)$ winding global symmetry $\tilde{\theta }\to \tilde{\theta }+c^w$ \cite{ref14} in the dual phase field $\tilde{\theta }$ is obtained as follows: 
\begin{eqnarray}\label{eq10}
\scalebox{0.90}{$\displaystyle\hspace{40px}
J_i^w\left( \mathbf{x},\tau  \right)\equiv \frac{\varepsilon_{ij}^{xy}}{2\pi }\left[ \nabla _j^{xy}\theta \left( \mathbf{x},\tau  \right)-2\pi n_j^{xy}\left( \mathbf{x},\tau  \right) \right]
$},
 \end{eqnarray}
From the partition function on the right-hand side of Eq. (\ref{eq3}), the dual Lagrangian on the Lagrangian in Eq. (\ref{eq8}) follows. 

\begin{eqnarray}\label{eq11}
\scalebox{0.9}{$\displaystyle\hspace{20px}
\tilde{L}\!=\!-\frac{( \tilde{\beta' }_i^{xy} )_v}{2}\!\!\sum\limits_{dual-plaquette}\!\left( \nabla_i^{xy}\tilde{\theta }\!-\!2\pi \tilde{n}_i^{xy} \right)^2\!-\!i\sum\limits_{dual-cube}{\varepsilon_{ij}^{xy}\tilde{n}_i^{xy}\nabla _j^{xy}\theta }
$},
 \end{eqnarray}
 where $\tilde{\beta }_0^{xy}$ and $\tilde{\beta }_1^{xy}$ are defined by $\tilde{\beta }_0^{xy}\equiv {\tilde{\beta }_0}$ and $\tilde{\beta }_1^{xy}\equiv {\tilde{\beta }_{xy}}$, respectively, and $\sum\limits_{dual-link}$, $\sum\limits_{dual-plaquette}$ and $\sum\limits_{dual-cube}$ are the sum of $dual\!-\!link$, $dual\!-\!plaquette$ and $dual-cube$ in subspace of $1+2d$ space, respectively. Following the same method as in Eq. (\ref{eq8}) through Eq. (\ref{eq10}), its dual momentum current $\tilde{J}_i^{xy}$ and dual winding current $\tilde{J}_i^w$ are, respectively, as follows: 
\begin{eqnarray}\label{eq12}
\scalebox{0.9}{$\displaystyle\hspace{60px}
\tilde{J}_i^{xy}( \mathbf{x},\tau )=-i{( \tilde{\beta' }_i^{xy} )}_v( \nabla _i^{xy}\tilde{\theta }-2\pi \tilde{n}_i^{xy} )\left( \mathbf{x},\tau  \right)
$},
 \end{eqnarray}
\begin{eqnarray}\label{eq13}
\scalebox{0.9}{$\displaystyle\hspace{60px}
\tilde{J}_i^w\left( \mathbf{x},\tau  \right)\equiv \frac{{\varepsilon }_{ij}}{2\pi }\left[ \nabla_j^{xy}\tilde{\theta }\left( \mathbf{x},\tau  \right)-2\pi \tilde{n}_j^{xy}\left( \mathbf{x},\tau  \right) \right]
$},
 \end{eqnarray}
\section{Dual transformation by the SVF for $XY$-plaquette model}
In this section, we apply the SVF \cite{ref7,ref8,ref9,ref10} of the dual transformation for the $1+1d$ XY-model \cite{ref22} shown in Appendix D to the case of the XY-plaquette model in $2+1d$.
The starting point is the partition function in the standard Villain approximation in Eq. (\ref{eq2}). Using Poisson's summation formula \cite{ref9,ref10} in Eq. (\ref{eqD1}) in Appendix D for Eq. (\ref{eq2}), as follows:
\begin{eqnarray}\label{eq14}
\scalebox{0.8}{$\displaystyle
Z_v\left( {\beta }_0,{\beta }_{xy} \right)\equiv C_{QV}\sum\limits_{\left\{ b \right\}}{\delta }_{\nabla b,0}\exp\sum\limits_{x,y,\tau }{\left\{ \frac{-b_{\tau }^2\left( \mathbf{x},\tau  \right)}{2\left( {\beta'}_0 \right)_v}+\frac{-b_{xy}^{2}\left( \mathbf{x},\tau  \right)}{2\left( {\beta'}_{xy} \right)_v} \right\}}
$},
 \end{eqnarray}
where $C_v\equiv \left[I_0\left(\beta'_0 \right)I_0\left(\beta'_{xy} \right) \right]^{M_{\tau }M_x}$ is the normalization parameter, and $b_i\left( \mathbf{x},\tau  \right)$$\left( i=\tau ,xy \right)$ are integer value auxiliary magnetic fields \cite{ref10,ref22} where the condition $\nabla b\equiv{\nabla }_{\tau }b_{\tau }-{\nabla }_{xy}b_{xy}=0$ is satisfied by the part of the $\delta$-function. We introduce dual integer value fields $\tilde{b}_i\left( \mathbf{x},\tau  \right)$$\left( i=\tau ,xy \right)$ \cite{ref22} to $b_i\left( \mathbf{x},\tau  \right)$ as follows:
\begin{eqnarray}\label{eq15}
\scalebox{1.0}{$\displaystyle\hspace{60px}
b_{i}^{xy}\left( x,y,\tau  \right)\equiv \varepsilon _{ij}^{xy}\tilde{b}_{j}^{xy}\left( x,y,\tau  \right)
$},
\end{eqnarray}
where $b_0^{xy}$, $b_1^{xy}$, $\tilde{b}_0^{xy}$, and $\tilde{b}_1^{xy}$ are defined by $b_{0}^{xy}\equiv b_{\tau }$, $b_1^{xy}\equiv b_{xy}$, $\tilde{b}_0^{xy}\equiv\tilde{b}_{\tau }$, and $\tilde{b}_1^{xy}\equiv\tilde{b}_{xy}$ respectively, By using the dual transformations of Eq. (\ref{eq15}), the following equations are obtained for Eq. (\ref{eq14}):
\begin{eqnarray}\label{eq16}
\scalebox{0.8}{$\displaystyle
Z_v\left( \beta_0,\beta_{xy} \right)\equiv C_{QV}\sum\limits_{\left\{ \tilde{b} \right\}}\delta_{\varepsilon_{ij}\nabla_i\tilde{b}_j,0}\exp\sum\limits_{x,y,\tau }{\left\{ \frac{-{\tilde{b}_{\tau }}^2\left( x,y,\tau  \right)}{2{{\left( \beta'_{xy} \right)}_v}}+\frac{-{\tilde{b}_{xy}}^2\left( x,y,\tau  \right)}{2\left( \beta' _0 \right)}_v \right\}}
$},
\end{eqnarray}
Following Poisson's formula in Eq. (\ref{eqD5}) in Appendix D, rewriting the integer-valued field $\tilde{b}_{\mu }\left( x,y,\tau \right)$ into a continuous value field $\tilde{B}_i\left( x,y,\tau\right)$$\left( i=\tau,xy\right) $\cite{ref22}, as follows:
\begin{eqnarray}\label{eq17}
\scalebox{0.8}{$\displaystyle\hspace{-40px}
Z_v\left( \beta_0,\beta_{xy} \right)=C_{QV}\sum\limits_{\left\{ l \right\}}\delta_{\varepsilon_{ij}\nabla_i{l_j},0}\int\mathcal{D}\tilde{B}\exp \sum\limits_{x,y,\tau }{\left\{ \frac{-{\tilde{B}_0}^2}{2{\left(\beta'_{xy}\right)}_v}+\frac{-{\tilde{B}_{xy}}^2}{2\left(\beta'_0 \right)_v}+i2\pi l_{xy}\tilde{B}_0-i2\pi l_{\tau }\tilde{B}_{xy} \right\}}
$},
\end{eqnarray}
where $l_i\left( x,\tau  \right)$$\left( i=\tau ,xy \right)$ corresponds to the integer-valued current. Integrating over the continuous value field ${{\tilde{B}}_{i}}\left( x,y,\tau  \right)$ of Eq. (\ref{eq17}) yield the following equation:
\begin{eqnarray}\label{eq18}
\scalebox{0.8}{$\displaystyle
Z_v=R_{QV}\sum\limits_{\left\{ l \right\}}\delta_{\varepsilon_{ij}{\nabla }_i{l_j},0}\exp \left( -1 \right)\sum\limits_{x,y,\tau }\left\{ 2{\pi }^2\left(\beta'_0 \right)_v{l_{\tau }}^2\left( x,y,\tau  \right)+2{\pi }^2{\left(\beta'_{xy} \right)}_v{l_{xy}}^2\left( x,y,\tau  \right) \right\}
$},
\end{eqnarray}
The Kronecker delta, when rewritten in the integral form, allow the equations to be written as follows: 
\begin{eqnarray}\label{eq19}
\scalebox{0.8}{$\displaystyle
Z_v=R_{QV}\sum\limits_{\left\{ l \right\}}\int\mathcal{D}\tilde\theta\exp \left( -1 \right)\sum\limits_{x,y,\tau }\left\{ 2{\pi }^2\left( \beta'_0 \right)_vl_{\tau}^2+2{\pi }^2\left(\beta'_{xy} \right)_v{l_{xy}}^2-i\tilde{\theta }\omega'_{\tau xy} \right\}
$}.
\end{eqnarray}
where $\omega'_{\tau xy}$ is defined here as the curl of $l_i^{xy}$, following Eq. (\ref{eq4}), as follows:

\begin{eqnarray}\label{eq20}
\scalebox{1.0}{$\displaystyle\hspace{60px}
\omega'_{\tau xy}\left( x,y,\tau  \right)\equiv \varepsilon _{ij}^{xy}\nabla _i^{xy}l_j^{xy}\left( x,y,\tau  \right)
$}.
\end{eqnarray}
where the integer-valued currents $l_0^{xy}$ and $l_1^{xy}$ are defined by $l_0^{xy}\equiv l_{\tau }$ and $l_1^{xy}\equiv l_{xy}$, respectively. 
Using the identity in Eq. (\ref{eqD1}) in Eq. (\ref{eq19}), we obtain the following form \cite{ref22}, 
\begin{eqnarray}\label{eq21}
\scalebox{0.9}{$\displaystyle
Z_v=C_v\sum\limits_{\left\{ \tilde{n}\right\}}\int\mathcal{D}\tilde{\theta }\exp \sum\limits_{x,y,\tau }{\left\{ \frac{-{\left( {\nabla }_{\tau }\tilde{\theta }-2\pi \tilde{n}_0 \right)}^2}{8{\pi }^2\left( \beta'_{xy} \right)_v}-\frac{\left( {\nabla }_{xy}\tilde{\theta }-2\pi\tilde{n}_{xy} \right)^2}{8{\pi }^2\left( \beta'_0\right)_v} \right\}}
$}.
\end{eqnarray}
Imposing the same relation ${(\tilde{\beta'}_0 )}_v=1/4{\pi }^2(\beta'_{xy} )_v$ and ${( \tilde{\beta'}_{xy} )}_v=1/4{\pi }^2(\beta'_0 )_v$ as in Eq. (\ref{eq6}) and (\ref{eq7}) on Eq. (\ref{eq21}), the self duality relation is established as in Eq. (\ref{eq3}), as follows:

\begin{eqnarray}\label{eq22}
\scalebox{0.8}{$\displaystyle\hspace{-70px}
Z_v\left( \beta_0,\beta_{xy} \right)\equiv R_{QV}\int\mathcal{D}\theta\sum\limits_{\left\{ n_{\tau },n_{xy} \right\}}\exp\sum\limits_{x,y,\tau }{\left\{ \frac{-\left( \beta'_0 \right)_v}{2}{{\left({\nabla }_{\tau }\theta -2\pi n_{\tau }\right)}^2}-\frac{\left( \beta'_{xy} \right)_v}{2}\left({\nabla }_{xy}\theta -2\pi n_{xy}\right)^2 \right\}}
$}\nonumber\\
 \scalebox{0.8}{$\displaystyle\hspace{-70px}
=C_v\sum\limits_{\left\{ \tilde{n} \right\}}\int\mathcal{D}\tilde{\theta }\exp \sum\limits_{x,y,\tau }{\left\{ \frac{-( \tilde\beta'_0)_v}{2}( {\nabla }_{\tau }\tilde{\theta }-2\pi \tilde{n}_0)^2-\frac{( \tilde\beta'_{xy})_v}{2}( \nabla_{xy}\tilde{\theta }-2\pi\tilde{n}_{xy} )^2\right\}}\equiv C_v{{\tilde{R}}^{-1}}_{QV}\tilde{Z}_V( \tilde{\beta }_0,\tilde{\beta }_x)
$},
\end{eqnarray}
\begin{eqnarray}\label{eq23}
\scalebox{0.85}{$\displaystyle\hspace{-70px}
\tilde{Z}_V( {\tilde{\beta }_0},{\beta}_x)\equiv\tilde{R}_{QV}\sum\limits_{\left\{\tilde{n}\right\}}\int{\mathcal{D}\tilde{\theta}}\exp\sum\limits_{x,y,\tau}{\left\{\frac{-(\tilde{\beta'}_0)_v}{2}\left({\nabla}_{\tau}\tilde{\theta}-2\pi\tilde{n}_0\right)^2-\frac{(\tilde{\beta'}_{xy})_v}{2}\left({\nabla}_{xy}\tilde{\theta}-2\pi\tilde{n}_{xy}\right)^2\right\}}
 $},\nonumber\\
\scalebox{0.85}{$\displaystyle\hspace{80px}
\tilde{R}_{QV}\equiv {\left[ R_v( \tilde{\beta'}_0 )R_v( \tilde{\beta'}_{xy}) \right]}^{M_{\tau}M_{xy}}
 $}.
\end{eqnarray}
The only difference with the self duality of the Modified Villain formulation in Eq. (\ref{eq3}) is the presence of an additional term  +$i\omega\tilde{\theta }$ . 
\section{Dual Hamiltonian method of the XY-plaquette model}
In this section, we show that it is possible to construct an exact self dual system for the Villain approximate version of the XY plaquette model without using Poisson's summation formula by implementing the algorithm of the DHM \cite{ref22,ref23} shown in Appendix C.
\subsection{Hamilton's canonical equations for the XY-plaquette model}
In the partition function of the Villain approximation of Eq. (\ref{eq2}), introducing an auxiliary field $N_{\theta }$ for its imaginary time component yields:
\begin{eqnarray}\label{eq24}
\scalebox{0.82}{$\displaystyle\hspace{-80px}
Z_v\left( {\beta }_0,{\beta }_{xy} \right)\!=\!R_{QV}\!\!\int\!\!\mathcal{D}\!{N}_{\theta }\!\!\int\mathcal{D}\theta\!\!\sum\limits_{\left\{ n_{\tau },n_{xy} \right\}}\!\!\exp\!\sum\limits_{x,y,\tau }{\left\{ i\left( {\nabla }_{\tau}\theta\! -\!2\pi n_{\tau }\right)N_{\theta }\!-\!E'_{pc}{N_{\theta }}^2\!-\!\frac{\left( \beta'_{xy} \right)_v}{2}{\left({\nabla }_{xy}\theta \!-\!2\pi n_{xy} \right)}^2 \right\}}
 $},
\end{eqnarray}
\begin{eqnarray}\label{eq25}
\scalebox{0.95}{$\displaystyle\hspace{90px}
E_{pc}\equiv \frac{{\hbar }^2}{2{\left( {{\beta }_0} \right)_v}{{\left( \Delta \tau  \right)}^2}}
 $},
\end{eqnarray}
where ${N}_{\theta }$ and ${E}_{pc}\equiv {E'}_{pc}{\hbar }/{\Delta \tau }$ are the winding number corresponding to the particle number, and the pseudo charging energy, respectively, and are interpreted as analogous to the particle number of Cooper pair, and the charging energy of the Josephson junction described, respectively in Eq. (\ref{eqC1}) in Appendix C. Therefore, from the Legendre transformation, the Euclidean Lagrangian $L_{\theta }$ and Hamiltonian $H_{\theta }$ are as follows, respectively: 
\begin{eqnarray}\label{eq26}
\scalebox{0.95}{$\displaystyle
L_{\theta}\left(\nabla_{\tau}\theta ,\theta \right)\equiv\frac{i}{\Delta \tau }\sum\limits_{x,y}{\left({\nabla }_{\tau}\theta-2\pi{{n}{\tau}}\right)}p_{\theta}+H_{\theta }\left(\theta ,N\right) 
$}
\end{eqnarray}
\begin{eqnarray}\label{eq27}
\scalebox{0.95}{$\displaystyle
H_{\theta }\left( \theta ,N \right)\equiv \frac{E_{pc}}{{\hbar }^2}\sum\limits_{x,y}p_{\theta}^2+\frac{{{\left( {{\beta }_{xy}} \right)}_v}}{2}\sum\limits_{x,y}{{{\left({\nabla }_{xy}\theta -2\pi n_{xy} \right)}^2}}
 $},
\end{eqnarray}
where $p_{\theta }\equiv \hbar N_{\theta }$ is the momentum of canonical conjugate to ${\nabla }_{xy}\theta'\equiv {\nabla }_{xy}\theta -2\pi n_{xy}$ and is defined as follows: 
\begin{eqnarray}\label{eq28}
\scalebox{0.95}{$\displaystyle\hspace{80px}
p_{\theta }\left( \mathbf{x},\tau  \right)\equiv i\Delta \tau \frac{\partial L_{\theta}\left(\nabla_{xy}\theta ,\nabla_{\tau}\nabla_{xy}\theta \right)}{\partial \left( \nabla_{\tau }\theta -2\pi n_{\tau } \right)}
 $}.
\end{eqnarray}
The Poisson bracket $\left\{...,...\right\}_p$  between $\nabla_{xy}\theta'\equiv{\nabla }_{xy}\theta -2\pi n_{xy}$ and $p_{\theta }\equiv \hbar N_{\theta }$ is as follows: 
\begin{eqnarray}\label{eq29}
\scalebox{1.0}{$\displaystyle\hspace{60px}
\left\{ {\nabla }_{xy}{\theta' }\left( \mathbf{x},\tau  \right),\hbar N_{\theta }\left( \mathbf{x'},\tau\right) \right\}_P=i\hbar {\delta }^D\left( \mathbf{x}-\mathbf{x'} \right)
 $},
\end{eqnarray}
From Eq. (\ref{eq27}), Hamilton's canonical equations are as follows \cite{ref22,ref23}:
\begin{eqnarray}\label{eq30}
\scalebox{0.95}{$\displaystyle\hspace{90px}
i\frac{\partial\nabla_{xy}\theta'\left( \mathbf{x},\tau\right)}{\partial \tau }=\frac{2E_{pc}}{\hbar }{N_{\theta }}\left( \mathbf{x},\tau\right)
 $},
\end{eqnarray}
\begin{eqnarray}\label{eq31}
\scalebox{0.95}{$\displaystyle\hspace{90px}
i\hbar \frac{\partial N_{\theta }\left( \mathbf{x},\tau  \right)}{\partial \tau }=-{\left( \beta _{xy} \right)_v}\nabla_{xy}\theta'\left( \mathbf{x},\tau  \right)
 $},
\end{eqnarray}
\subsection{Hamilton's canonical equations for the dual XY-plaquette model}
 On the other hand, the dual XY-plaquette model and its Villain approximation, which are dual to the partition function in Eqs. (\ref{eq1}) and (\ref{eq2}), respectively, are as follows: 
\begin{eqnarray}\label{eq32}
\scalebox{0.80}{$\displaystyle
\tilde{Z}\left( \tilde{\beta }_0,\tilde\beta_{xy} \right)=\int\mathcal{D}\tilde{\theta }\exp \sum\limits_{x,y,\tau }{\left[ -\tilde{\beta'}_0\left( 1-\cos{\nabla }_{\tau}\tilde{\theta } \right)-{\tilde{\beta'}_{xy}}\left( 1-\cos{\nabla }_{xy}\tilde{\theta } \right) \right]}
 $},
\end{eqnarray} 
\begin{eqnarray}\label{eq33}
\scalebox{0.8}{$\displaystyle\hspace{-40px}
\tilde{Z}_v\left( \tilde{\beta }_0,\tilde\beta_{xy}\right)\!\equiv\!\tilde{R}_{QV}\!\!\int\!\mathcal{D}\tilde{\theta }\!\!\sum\limits_{\left\{ \tilde{n}_\tau,\tilde{n}_{xy} \right\}}\!\!\!\exp\!\sum\limits_{x,y,\tau }{\left\{ \frac{-(\tilde\beta'_0 )_v}{2}{\left({\nabla }_{\tau }\tilde{\theta }\!-\!2\pi \tilde{n}_{\tau }\right)^2}\!-\!\frac{{( \tilde\beta'_{xy} )}_v}{2}\left(\nabla_{xy}\tilde{\theta }-2\pi \tilde{n}_{xy} \right)^2 \right\}}
 $},
\end{eqnarray} 
Similar to Eq. (\ref{eq24}), introducing the auxiliary field $\tilde{N}_{\theta }$ for the imaginary time component of Eq. (\ref{eq33}), as follows: 
\begin{eqnarray}\label{eq34}
\scalebox{0.8}{$\displaystyle\hspace{-80px}
\tilde{Z}_v\left( \tilde{\beta }_0,\tilde{\beta }_{xy} \right)\!=\!\tilde{R}_{QV}\!\!\int\!\!\mathcal{D}\tilde{N}_{\tilde{\theta }}\!\!\int\!\!\mathcal{D}\tilde{\theta }\!\!\!\sum\limits_{\left\{ \tilde{n}_{\tau },\tilde{n}_{xy} \right\}}\!\!\!\!\exp\!\sum\limits_{x,y,\tau }{\left\{ i\left( \nabla_{\tau }\tilde{\theta }\!-\!2\pi {\tilde{n}_{\tau }} \right)\tilde{N}_{\tilde{\theta }}\!-\!E'_{pi}\tilde{N}_{\tilde{\theta}}^2\!-\!\frac{\left( {\beta'}_{xy} \right)_v}{2}\left( {\nabla }_{xy}\theta\! -\!2\pi{n}_{xy} \right)^2 \right\}}
$},
\end{eqnarray} 
\begin{eqnarray}\label{eq35}
\scalebox{0.95}{$\displaystyle\hspace{80px}
E_{pi}\equiv \frac{{\hbar }^2}{2{( \tilde{\beta }_0 )_v}{{\left( \Delta \tau  \right)}^2}}
$},
\end{eqnarray} 
where $\tilde{N}_{\tilde{\theta }}$ and $E_{pi}\equiv {E'}_{pi}{\hbar }/{\Delta \tau }\;$are the dual winding number in a dual relation to $N_{\theta }$ in Eq. (\ref{eq24}) and the pseudo inductance energy, respectively, and are interpreted as analogies to the particle number of magnetic flux-quantum and the inductance energy of the QPS junction, respectively, described in Eq. (\ref{eqC3}) of Appendix C. Thus, the Euclidean dual Lagrangian and dual Hamiltonian are as follows, respectively: 
\begin{eqnarray}\label{eq36}
\scalebox{0.9}{$\displaystyle
\tilde{L}_{\tilde{\theta }}\left(\nabla_{xy}\tilde{\theta },\nabla_{\tau }\nabla_{xy}\tilde{\theta } \right)\equiv -\frac{i}{\Delta \tau }\sum\limits_{x,y}{\left( {{\nabla }_{\tau }}\tilde{\theta }-2\pi \tilde{n}_{\tau } \right)}\tilde{p}_{\tilde{\theta }}\left( x,\tau  \right)+\tilde{H}_{\tilde{\theta }}\left( \tilde{\theta },\tilde{N} \right)
 $},
\end{eqnarray}
\begin{eqnarray}\label{eq37}
\scalebox{0.9}{$\displaystyle
\tilde{H}_{\tilde{\theta }}\left({\nabla }_{xy}\tilde{\theta },{\tilde{p}}_{\tilde{\theta }} \right)\equiv \frac{E_{pi}}{{\hbar }^2}\sum\limits_{x,y}{\tilde{p}_{\tilde{\theta}}^2\left( x,\tau  \right)}+\frac{( \tilde{\beta }_{xy} )_v}{2}\sum\limits_{x,y}\left( {\nabla }_{xy}\tilde{\theta }-2\pi \tilde{n}_{xy} \right)^2
$}.
\end{eqnarray}
where $\tilde{p}_{\tilde{\theta }}\equiv \hbar \tilde{N}_{\tilde{\theta }}$ is the canonical conjugate momentum to ${\nabla }_{xy}\tilde{\theta }$ and is defined as follows. 
\begin{eqnarray}\label{eq38}
\scalebox{0.9}{$\displaystyle\hspace{80px}
\tilde{p}_{\tilde{\theta }}\left( \mathbf{x},\tau  \right)\equiv i\Delta \tau \frac{\partial\tilde{L}_{\tilde{\theta}}\left({\nabla }_{xy}\tilde{\theta },{{\nabla }_{\tau }}{{\nabla }_{xy}}\tilde{\theta } \right)}{\partial \left({\nabla }_{\tau }\tilde{\theta }-2\pi {\tilde{n}}_{\tau } \right)}
$}.
\end{eqnarray}
 The Poisson bracket between $\nabla_{xy}\tilde{\theta'}\equiv \nabla_{xy}\tilde{\theta }-2\pi \tilde{n}_{xy}$ and $\tilde{p}_{\tilde{\theta }}\equiv \hbar \tilde{N}_{\tilde{\theta}}$ is as follows:
\begin{eqnarray}\label{eq39}
\scalebox{0.9}{$\displaystyle\hspace{60px}
\left\{ \nabla_{xy}\tilde{\theta' }\left( \mathbf{x},\tau  \right),\hbar \tilde{N}_{\tilde{\theta }}\left( \mathbf{x'},\tau  \right) \right\}_P=i\hbar {\delta }^D\left( \mathbf{x}-\mathbf{x'} \right)
$}.
\end{eqnarray}
From Eq. (\ref{eq37}), Hamilton's canonical equations are as follows:
\begin{eqnarray}\label{eq40}
\scalebox{0.95}{$\displaystyle\hspace{100px}
i\frac{\partial\nabla_{xy}\tilde{\theta'}\left( \mathbf{x},\tau\right)}{\partial\tau }=\frac{2E_{pi}}{\hbar }{\tilde{N}_{\tilde{\theta }}}\left( \mathbf{x},\tau\right)
$},
\end{eqnarray}
 \begin{eqnarray}\label{eq41}
\scalebox{0.95}{$\displaystyle\hspace{100px}
i\hbar \frac{\partial \tilde{N}_{\tilde{\theta}}\left( \mathbf{x},\tau\right)}{\partial \tau }=-( \tilde{\beta }_{xy})_v\partial\nabla_{xy}\tilde{\theta'}\left( \mathbf{x},\tau\right)
$}.
\end{eqnarray}
\subsection{Pseudo-Josephson equation and self-dual conditions for XY-plaquette and dual XY-plaquette models}
By analogy with the Josephson equation for superconductivity, Eq. (\ref{eq30}) and (\ref{eq31}) can be rewritten as the following pseudo-Josephson equations, 
\begin{eqnarray}\label{eq42}
\scalebox{0.9}{$\displaystyle\hspace{60px}
V_{xy}\equiv \frac{i\hbar }{2e}\frac{\partial{\nabla_{xy}}{\theta' }\left( \mathbf{x},\tau\right)}{\partial \tau }=\frac{2E_{pc}}{2e}{N_{\theta }}\left( \mathbf{x},\tau  \right)
$},
\end{eqnarray}
 \begin{eqnarray}\label{eq43}
\scalebox{0.9}{$\displaystyle\hspace{60px}
I_{xy}\equiv i\left( 2e \right)\frac{\partial{N_{\theta }}\left( \mathbf{x},\tau  \right)}{\partial \tau }=-\frac{2\pi }{\Phi_0}{\left(\beta_{xy} \right)_v}{\nabla_{xy}}{\theta' }\left( \mathbf{x},\tau\right)
$},
\end{eqnarray}
where $V_{xy}$ and $I_{xy}$ are the pseudo-Josephson voltage and pseudo-Josephson current in the $XY$-plaquette model, respectively. Similarly, Eq. (\ref{eq40}) and (\ref{eq41}) can be rewritten as the following dual pseudo-Josephson equations,
\begin{eqnarray}\label{eq44}
\scalebox{0.9}{$\displaystyle\hspace{60px}
\tilde{V}_{xy}\equiv \frac{i\hbar }{\Phi_0}\frac{\partial{\nabla }_{xy}\tilde{\theta'}\left( \mathbf{x},\tau  \right)}{\partial \tau }=\frac{2E_{pi}}{\Phi_0}{{\tilde{N}}_{{\tilde{\theta }}}}\left( \mathbf{x},\tau  \right) 
$},
\end{eqnarray}
 \begin{eqnarray}\label{eq45}
\scalebox{0.9}{$\displaystyle\hspace{60px}
\tilde{I}_{xy}=-i{\Phi }_0\frac{\partial{\tilde{N}_{\tilde{\theta}}}\left( \mathbf{x},\tau  \right)}{\partial \tau }=\Phi_0\frac{(\tilde{\beta}_{xy})_v}{\hbar }{\nabla }_{xy}\tilde{\theta'}\left( \mathbf{x},\tau  \right)
$},
\end{eqnarray}
Where $\tilde{V}_{xy}$ and $\tilde{I}_{xy}$ are the pseudo-dual Josephson voltage and pseudo-dual Josephson current \cite{ref22,ref23} in the $XY$-plaquette model, respectively. The following self dual conditions are imposed between the pseudo-Josephson equations of Eq. (\ref{eq42}) and (\ref{eq43}), and the pseudo-dual Josephson equations of Eq. (\ref{eq44}) and (\ref{eq45}) according to algorithm of the DHM \cite{ref22,ref23} as follows:\\
\begin{enumerate}
      \item $V_{xy}=\tilde{I}_{xy}$: Pseudo Josephson voltage equals pseudo dual Josephson current.
\begin{eqnarray}\label{eq46}
\scalebox{0.9}{$\displaystyle\hspace{60px}
\frac{i\hbar }{2e}\frac{\partial{\nabla }_{xy}\theta'\left( \mathbf{x},\tau\right)}{\partial\tau}=-i{\Phi }_0\frac{\partial\tilde{N}_{\tilde{\theta }}\left( \mathbf{x},\tau  \right)}{\partial \tau}
$},
\end{eqnarray}
 \begin{eqnarray}\label{eq47}
\scalebox{0.9}{$\displaystyle\hspace{60px}
\frac{2E_{pc}}{2e}{N_{\theta }}\left( \mathbf{x},\tau\right)=\Phi_0\frac{(\tilde{\beta }_{xy})_v}{\hbar }\nabla_{xy}\tilde{\theta'}\left( \mathbf{x},\tau  \right)
$},
\end{eqnarray}
 \item $I_{xy}=\tilde{V}_{xy}$: Pseudo Josephson current equals pseudo dual Josephson voltage.
\begin{eqnarray}\label{eq48}
\scalebox{0.9}{$\displaystyle\hspace{60px}
i\left(2e\right)\frac{\partial N_{\theta }\left( \mathbf{x},\tau  \right)}{\partial\tau}=\frac{i\hbar }{\Phi_0}\frac{\partial\nabla_{xy}\tilde{\theta'}\left( \mathbf{x},\tau  \right)}{\partial\tau}
$},
\end{eqnarray}
 \begin{eqnarray}\label{eq49}
\scalebox{0.9}{$\displaystyle\hspace{60px}
-\frac{2\pi\left( \beta_{xy} \right)_v}{\Phi_0}\nabla_{xy}\theta' \left( \mathbf{x},\tau\right)=\frac{2E_{pi}}{\Phi_0}\tilde{N}_{\tilde{\theta}}\left( \mathbf{x},\tau  \right)
$},
\end{eqnarray}
\end{enumerate}
First, integrating both sides of Eq. (\ref{eq46}), one of the self dual conditions in i), we obtain the following relational expression. 
 \begin{eqnarray}\label{eq50}
\scalebox{0.9}{$\displaystyle\hspace{60px}
\tilde{N}_{\tilde{\theta }}\left( \mathbf{x},\tau\right)=-\frac{1}{2\pi }\Bigl[\nabla_{xy}\theta \left( \mathbf{x},\tau  \right)-2\pi n_{xy}\left( \mathbf{x},\tau  \right) \Bigr]
$},
\end{eqnarray}
From Eq. (\ref{eq47}), which is another self-dual condition of i), solving for $\hbar N$ yields the following.
 \begin{eqnarray}\label{eq51}
\scalebox{0.9}{$\displaystyle\hspace{60px}
N_{\theta }\left( \mathbf{x},\tau  \right)=\frac{2\pi(\tilde{\beta }_{xy})_v}{2E_{pc}}\left[ {{\nabla }_{xy}}\tilde{\theta }\left( \mathbf{x},\tau  \right)-2\pi \tilde{n}_{xy}\left( \mathbf{x},\tau\right) \right]
$},
\end{eqnarray}
Next, integrating both sides of Eq. (\ref{eq48}), one of the self dual conditions in ii), we obtain the following relational expression. 
 \begin{eqnarray}\label{eq52}
\scalebox{0.9}{$\displaystyle\hspace{60px}
N_{\theta }\left( \mathbf{x},\tau \right)=\frac{1}{2\pi }\left[\nabla_{xy}\tilde{\theta }\left( \mathbf{x},\tau\right)-2\pi\tilde{n}_{xy}\left( \mathbf{x},\tau\right) \right]
$},
\end{eqnarray}
From Eq. (\ref{eq49}), which is another self-dual condition of ii), solving for ${{\tilde{N}}_{\theta }}$ yields the following.
 \begin{eqnarray}\label{eq53}
\scalebox{0.9}{$\displaystyle\hspace{60px}
\tilde{N}_{\tilde{\theta }}\left( \mathbf{x},\tau \right)=-\frac{2\pi {{\left( {{\beta }_{xy}} \right)}_{v}}}{2{{E}_{pi}}}\left[ {{\nabla }_{xy}}\theta \left( \mathbf{x},\tau  \right)-2\pi {{n}_{xy}}\left( \mathbf{x},\tau  \right) \right]
$},
\end{eqnarray}
Comparing Eq. (\ref{eq50}) and (\ref{eq53}), the following relationship holds:
 \begin{eqnarray}\label{eq54}
\scalebox{1.0}{$\displaystyle\hspace{100px}
E_{pi}=2{\pi}^2\left({\beta}_{xy} \right)_v
$},
\end{eqnarray}
Similarly, comparing Eq. (\ref{eq51}) and (\ref{eq52}), the following relation holds: 
 \begin{eqnarray}\label{eq55}
\scalebox{1.0}{$\displaystyle\hspace{100px}
E_{pc}=2{{\pi }^2}(\tilde{\beta }_{xy})_v
$},
\end{eqnarray}
From the relationship between Eq. (\ref{eq25}) and (\ref{eq55}), and between Eq. (\ref{eq35}) and (\ref{eq54}), the following relationship \cite{ref22,ref23} can be obtained, 
 \begin{eqnarray}\label{eq56}
\scalebox{1.0}{$\displaystyle\hspace{20px}
2{\pi}^2(\tilde{\beta}_{xy})_v=\frac{{\hbar}^2}{2\left(\beta_0 \right)_v\left(\Delta\tau\right)^2},\;2{\pi}^2\left(\beta_{xy}\right)_v=\frac{{\hbar}^2}{2( \tilde{\beta}_0)_v\left(\Delta\tau\right)^2}
$},
\end{eqnarray}
This relationship is consistent with Eq. (\ref{eq6}) and (\ref{eq7}). As shown below, From Eq. (\ref{eq50})$\sim$(\ref{eq55}) above, it follows that the winding number $N_{\theta}$ is equal to the dual winding current $\tilde{J}_0^w$ as defined in Eq. (\ref{eq13}),
 \begin{eqnarray}\label{eq57}
\scalebox{1.0}{$\displaystyle\hspace{100px}
N_{\theta }\left( \mathbf{x},\tau\right)=\tilde{J}_0^w\left( \mathbf{x},\tau\right)
$}.
\end{eqnarray}
And it follows that the number of dual windings $\tilde{N}_{\theta}$ is equal to the winding current $-J_0^w$ defined in Eq. (\ref{eq10}),
 \begin{eqnarray}\label{eq58}
\scalebox{1.0}{$\displaystyle\hspace{100px}
\tilde{N}_{\tilde{\theta }}\left(\mathbf{x},\tau \right)=-J_0^w\left( \mathbf{x},\tau  \right)
$}.
\end{eqnarray}
 On the other hand, from the variational method for $N_{\theta }$ in the Lagrangian of Eq. (\ref{eq26}), solving for $N_{\theta }$ as follows:
 \begin{eqnarray}\label{eq59}
\scalebox{1.0}{$\displaystyle\hspace{60px}
N_{\theta}\left(\mathbf{x},\tau \right)=i\left(\beta'_0 \right)_v\Bigl[\nabla_{\tau}\theta\left(\mathbf{x},\tau \right) -2\pi n_{\tau }\left(\mathbf{x},\tau \right)\Bigr]
$}.
\end{eqnarray}
As in Eq. (\ref{eq59}), from the variational method for $\tilde{N}_{\theta}$ in the Lagrangian of Eq. (\ref{eq36}), can solving for $\tilde{N}_{\theta }$ as follows :
 \begin{eqnarray}\label{eq60}
\scalebox{1.0}{$\displaystyle\hspace{60px}
\tilde{N}_{\theta}\left( \mathbf{x},\tau\right)=i(\tilde{\beta'_0})_v\left[\nabla_{\tau}\tilde{\theta}\left(\mathbf{x},\tau\right)-2\pi \tilde{n}_{\tau}\left( \mathbf{x},\tau\right) \right]
$}.
\end{eqnarray}
Since Eq. (\ref{eq57}) and (\ref{eq59}), and Eq. (\ref{eq58}) and (\ref{eq60}) are equal, $N_{\theta }$ and $\tilde{N}_{\tilde{\theta }}$ are as follows:
 \begin{eqnarray}\label{eq61}
\scalebox{1.0}{$\displaystyle\hspace{40px}
\frac{1}{2\pi}\left({\nabla }_{xy}\tilde{\theta }-2\pi \tilde{n}_{xy} \right)=i\left( \beta'_0\right)_v^2\Bigl({\nabla }_{\tau }\theta -2\pi n_{\tau } \Bigr)
$}.
\end{eqnarray}
 \begin{eqnarray}\label{eq62}
\scalebox{1.0}{$\displaystyle\hspace{40px}
\frac{-1}{2\pi}\Bigl(\nabla_{xy}\theta -2\pi n_{xy} \Bigr)=i( \tilde{\beta'_0} )_v\left({\nabla }_{\tau }\tilde{\theta }-2\pi {\tilde{n}_{\tau }} \right)
$}.
\end{eqnarray}

The relations in Eqs. (\ref{eq61}), (\ref{eq62}) and (\ref{eq56}) allow us to perform an exact dual transformation from Hamiltonian $H_{\theta}$ in Eq. (\ref{eq27}) to dual Hamiltonian $\tilde{H}_{\theta}$ in Eq. (\ref{eq37}), without using Poisson's summation formula.

\section{summary and discussion}
In Section 2, as an example of a self-dual system in a $1+1d$ similar system, we reviewed the dual transformation of the XY plaquette model in the $1+2d$ system according to the modified Villain formulation introduced by Gorantla \cite{ref14} et al. The greatest advantage of the dual transformation in the modified Villain formulation is its great simplicity, since the dual transformation can be performed directly by using Poisson's sum formula only once.In Section 3, we showed that the XY plaquette model is a strictly self dual system by performing a dual transformation in the usual Villain formulation, following the method of our previous paper. The advantage of this method is that it naturally shows that the XY plaquette model is strictly self dual, without adding an artificial extra term $+i\omega \tilde{\theta }$ as in Eqs. (\ref{eq3}) or (\ref{eqA8}), as in the case of the modified Villain formulation. The disadvantage is that it is very complicated, since the Poisson's sum formula must be used three times and then the path integral must also be used. In Section 4, the dual Hamiltonian method submitted in our previous paper \cite{ref22,ref23} was used between the XY plaquette model and the dual XY plaquette model to obtain the results in Eqs. (\ref{eq57}) $\sim$ (\ref{eq62}). Furthermore, these results can be summarized by the following two equations. 

 \begin{eqnarray}\label{eq63}
\scalebox{1.0}{$\displaystyle\hspace{80px}
J_i^w\left( \mathbf{x},\tau \right)=\sigma_{ij}^z\tilde{J}_j^{xy}( \mathbf{x},\tau )
$},
\end{eqnarray}
 \begin{eqnarray}\label{eq64}
\scalebox{1.0}{$\displaystyle\hspace{80px}
\tilde{J}_i^w\left( \mathbf{x},\tau  \right)=\sigma_{ij}^z J_j^{xy}( \mathbf{x},\tau )
$},
\end{eqnarray}
where $\sigma_{ij}^z$ is the $z$ component of the Pauli matrix satisfying $\sigma_{\tau,\tau}^z=-\sigma_{xy,xy}^z=1$ and $\sigma_{\tau,xy}^z=-\sigma_{xy,\tau}^z=0$. 
The left hand side of Eq. (\ref{eq63}) refers to the winding current in the original system, while the right hand side of Eq. (\ref{eq63}) represents the momentum current in the dual system. On the other hand, the left hand side of Eq. (\ref{eq64}) means the winding current in the dual system, and the right side of Eq. (\ref{eq64}) represents the momentum current in the original system. One of the greatest advantage of the dual Hamiltonian method is that it can directly perform the dual transformation, where it is self-dual, using the relational formulas in Eqs. (\ref{eq63}) and (\ref{eq64}), without using the Poisson sum formula or path integral at all. Another great advantage is that the relationship between the winding currents of the original system and the momentum currents of the dual system is clarified by the relational relations in Eqs. (\ref{eq63}) and (\ref{eq64}).

\appendix
\section{Dual transformation by MVF for $1+1d$ XY-model}
In this Appendix, we describe the dual transformation in the $1+1d$ XY-model, following the MVF by Sulejmanpasic et al. \cite{ref11} and Gorantl \cite{ref14} et al. As a starting point, the partition function of the lattice version of the $1+1d$ XY model is as follows:

 \begin{eqnarray}\label{eqA1}
\scalebox{1.0}{$\displaystyle
 Z\left( {\beta_0},{\beta_x} \right)=\int{\mathcal{D}\theta }\exp \sum\limits_{x,\tau }{\Biggl[ -{\beta }'_0\left( 1-\cos {\nabla_{\tau }}\theta  \right)-{\beta }'_x\left( 1-\cos {\nabla_x}\theta  \right) \Biggr]}
$},\nonumber \\
 \scalebox{0.8}{$\displaystyle
 \int{\mathcal{D}\theta}\equiv {\prod\limits_{\tau =0}^{N_\tau}\prod\limits_{x=1}^{N_x}{\int\limits_{-\pi }^{\pi }\frac{\displaystyle d{\theta}\left( x, \tau \right)}{\displaystyle 2\pi }}}
 $},\quad
  {\beta }'_0\equiv\frac{{\beta }_0\Delta \tau }{\hbar },\;\;{\beta }'_x\equiv\frac{{\beta }_x\Delta \tau }{\hbar },
\end{eqnarray}
where$\;{\nabla }_{\tau }\theta\!\left( x,\tau\right)\!\equiv\!\theta\left( x,\tau \right)\!-\!\theta\!\left( x,\tau\!-\!a\right)$ and ${\nabla }_{\tau }\theta\!\left( x,\tau  \right)\!\equiv\!\theta \left( x,\tau  \right)\!-\!\theta \left( x\!-\!a,\tau  \right)$ are the difference operators for the imaginary time and space components, respectively. The sums $\sum\limits_{x}{\equiv }\sum\nolimits_{x=1}^{M_x}{}$ and $\sum\limits_{\tau }{\equiv }\sum\nolimits_{\tau =1}^{M_{\tau }}{}$ of the sum $\sum\limits_{x,\tau }{}$ are given as lattice summations of integral $\int_{-L/2\;}^{L/2\;}{dx}$ and $\int_0^{\beta }{d\tau }$, respectively. Where $\Delta\tau\equiv{\tau }_{\max }/M_{\tau }\;$, ${\tau }_{\max }$, $M_{\tau }$, ${M_x}\equiv L/a\;$, $L$, and $a$ are minimum imaginary time interval, the maximum imaginary time, the time division number, the space division number, the length of the $x$-space and the lattice spacing, respectively. ${{\beta }_{0}}$ and ${{\beta }_{x}}$ are the energies of the imaginary time $\tau$ and the $x$-space component, respectively, and ${{\beta }'_0}\;$ and ${\beta }'_{x}\;$ are the respective dimensionless energies. In the $\exp \left( \cos \theta  \right)$ part of Eq. (\ref{eqA1}), we introduce the following ansatz as a periodic Gaussian approximation \cite{ref9,ref10}: 
\begin{eqnarray}\label{eqA2}
\scalebox{1.0}{$
\hspace{50px} e^{\alpha\cos\theta}\to R_v\left(\alpha\right)\sum\limits_{n=-\infty }^{\infty }{{e^{\frac{-{{\beta }_v}\left( \alpha  \right)}{2}{\left( \theta -2\pi n \right)}^2}}}
 $},
\end{eqnarray}
where $n$ is an integer value in $\left[-\infty ,\infty\right]$, Eq. (\ref{eqA2}) is called Villain approximation \cite{ref9,ref10}, and ${R_v}\left( \alpha  \right)$ and ${\beta }_v\left( \alpha  \right)$ are Villain's normalization factor and Villain's inverse temperature, respectively, and are defined as follows:
\begin{eqnarray}\label{eqA3}
\scalebox{1.0}{$
{R_v}\!\left( \alpha  \right)\equiv {I_0}\left( \alpha  \right){{\left[ 2\pi{\beta }_v\!\left(\alpha\right) \right]}^{\frac{1}{2}}},\;{{\beta }_v}\!\left( \alpha  \right)\!\equiv\! {{\left\{ -2\ln \left[ {I_1\left( \alpha  \right)}/{I_0\left( \alpha  \right)}\; \right] \right\}}^{-1}}
 $},
\end{eqnarray}
where $I_0\left(\alpha\right)$ and $I_1\left(\alpha\right)$ represent the modified Bessel functions of order zero and first, respectively. $R_v\!\left( \alpha  \right)$ and ${\beta }_v\!\left( \alpha  \right)$ are in the limit $\alpha {\to} \infty$ and $\alpha {\to} 0$, the large $\alpha$ expansion and small $\alpha$ expansion are, respectively \cite{ref9,ref10}, as follows: 
$\bullet\;$\leftline{For large limit, the large $\alpha$ expansion}
\begin{eqnarray}\label{eqA4}
\scalebox{1.0}{$
\hspace{-30px}{{\beta }_v}\!\left( \alpha\right)\!\approx\!\alpha\!\left( 1\!-\!\frac{1}{2\alpha}\!-\!\frac{5}{24{{\alpha }^2}}\!-\!\frac{7}{24{{\alpha }^3}}...\right),\;{R_v}\!\left(\alpha\right)\!\approx\!{e^{\alpha }}\!\left( 1\!-\!\frac{1}{8\alpha }\!-\!\frac{37}{384{{\alpha }^2}}\!-\!\frac{433}{384{{\alpha }^3}}...\right),
 $}
\end{eqnarray}
$\bullet\;$\leftline{For small limit, the small $\alpha$ expansion}
\begin{eqnarray}\label{eqA5}
\scalebox{1.0}{$
\hspace{-30px}{{\beta }_v}\!\left( \alpha  \right)\!\approx\!-\!{{\left[2\ln\!\left( \frac{\alpha }{2}\right)\!-\!\frac{{\alpha }^2}{4}\!+\!\frac{5{\alpha }^4}{192}... \right]}^{-1}}\!,\;R_v\!\left( \alpha\right)\!\approx\!{{\left[ 2\pi{\beta }_v\!\left(\alpha\right) \right]}^{ \frac{1}{2}}}\left(1\!+\!\frac{\alpha }{4}\!+\!\frac{{\alpha }^4}{64}... \right),
 $}
\end{eqnarray}
Thus, the Villain approximation is a good approximation in both the large $\alpha$ and small $\alpha$ limits. Apply the Villain approximation to $Z$ in Eq. (\ref{eqA1}) and let $Z_v$ be its partition function as follows:

 \begin{eqnarray}\label{eqA6}
\scalebox{0.9}{$\displaystyle
\hspace{-50px}Z_v\left( {\beta }_0,{{\beta }_x} \right)\!\equiv\!R_{QV}\!\int\!{\mathcal{D}\theta }\!\sum\limits_{\left\{ n \right\}}\exp\!\sum\limits_{x,\tau }\!{\left\{ \frac{{\!-\!\left( {\beta }'_0 \right)}_v}{2}{{\left({\nabla }_{\tau }\theta\!-\!2\pi n_{\tau } \right)}^2}\!+\!\frac{{\!-\!\left( {\beta }'_x  \right)}_v}{2}{{\left( {\nabla }_x\theta\!-\!2\pi n_x \right)}^2} \right\}}
$},\nonumber \\
 \scalebox{0.9}{$\displaystyle\hspace{70px}
   R_{QV}\equiv {{\left[ R_v\left({\beta }'_0 \right)R_v\left( {\beta }'_x \right) \right]}^{M_{\tau }M_x}}.
 $}
\end{eqnarray}
The summation symbols $\sum\limits_{\left\{ n \right\}}{\equiv }\sum\limits_{n_0\left( x,\tau  \right)=-\infty }^{\infty }{\sum\limits_{n_x\left( x,\tau  \right)=-\infty }^{\infty }}$, and $\sum\limits_{n_0\left( x,\tau  \right)=-\infty }^{\infty }$ are $\sum\limits_{{n_x}\left( x,\tau  \right)=-\infty }^{\infty }$ used for the integer-valued fields $n_0\left( x,\tau  \right)$ and $n_x\left( x,\tau  \right)$, respectively. 
The integer-valued field $n_i\left( i=\tau ,x \right)$ is called the vortex gauge field or jumping number field, and together with the phase field $\theta \left( x,\tau  \right)$ , it is a gauge transformation as follows:
\begin{eqnarray}\label{eqA7}
\scalebox{0.95}{$\displaystyle
\!\!\!\theta \left( x,\tau  \right)\to \theta \left( x,\tau  \right)+2\pi \gamma \left( x,\tau  \right), n_0\left( x,\tau  \right)\to n_0\left( x,\tau  \right)+{\nabla }_{\tau }\gamma \left( x,\tau  \right)
$},\nonumber \\
 \scalebox{0.95}{$\displaystyle\hspace{50px} 
n_x\left( x,\tau  \right)\to n_x\left( x,\tau  \right)+{\nabla }_x\gamma \left( x,\tau  \right)
 $},
\end{eqnarray}
where $\gamma \left( x \right)$ is an arbitrary integer value real field. The MVF by Gorantla \cite{ref14} et al. adds an extra additional term $+i\tilde{\theta }\omega$ of the dual phase field $\tilde{\theta }$ to Eq. (\ref{eqA6}), as follows:
 \begin{eqnarray}\label{eqA8}
\scalebox{0.80}{$\displaystyle
\hspace{-50px} Z_v\left( {\beta }_0,{{\beta }_x} \right)\!\equiv\!R_{QV}\!\!\!\int\!\! {\mathcal{D}\theta }\!\! \int\!\! {\mathcal{D}\tilde{\theta }}\!\sum\limits_{\left\{ n \right\}}\exp\!\!\left( -1 \right)\!\sum\limits_{x,\tau }\!{\left\{ \frac{{\!\left( {\beta }'_0 \right)}_v}{2}{{\left({\nabla }_{\tau }\theta\!-\!2\pi n_{\tau } \right)}^2}\!+\!\frac{{\!\left( {\beta }'_x  \right)}_v}{2}{{\left( {\nabla }_x\theta\!-\!2\pi n_x \right)}^2} \!+\!i\tilde{\theta }\omega \right\}}
$},
\end{eqnarray}
where $\theta$ and $\tilde{\theta }$ are phases that are dual relation to each other, and $\omega$ is defined as curl of $R_v\!\left( E \right)$, as follows:
 \begin{eqnarray}\label{eqA9}
\scalebox{1.0}{$\displaystyle
\hspace{100px} \omega \left( x,\tau  \right)\equiv {{\varepsilon }_{ij}}{{\nabla }_i}{n_j}\left( x,\tau  \right)
$},
\end{eqnarray}
Sulejmanpasic et al.\cite{ref11} and Gorantla \cite{ref14} et al. modified the Poisson's summation formula for the purpose of creating a self dual partition function as follows: 
 \begin{eqnarray}\label{eqA10}
\scalebox{0.8}{$\displaystyle
\hspace{-50px} \sum\limits_{n=-\infty }^{\infty }\!{\exp\! \left\{ \frac{-\alpha }{2}{{\left( \theta -2\pi n \right)}^2}\!+\!in\tilde{\theta } \right\}}\!=\!\frac{1}{\sqrt{2\pi \alpha }}\!\sum\limits_{\tilde{n}=-\infty }^{\infty }\!{\exp \!\left\{ \frac{-1}{2{{\left( 2\pi  \right)}^2}\alpha }{{\left( \tilde{\theta }-2\pi \tilde{n} \right)}^2}\!+\!\frac{i\theta }{2\pi }\left( \tilde{\theta }-2\pi \tilde{n} \right) \right\}}
$},
\end{eqnarray}
Eq. (\ref{eqA10}) is the starting point for the dual transformation by the Modified Villain formulation in one component. 
To more clearly express the self-duality of Eq. (\ref{eqA10}), we rewrite it as follows: 
 \begin{eqnarray}\label{eqA11}
\scalebox{0.9}{$\displaystyle
\hspace{20px} Z\left( \alpha  \right)=\frac{1}{\sqrt{2\pi \alpha }}\exp \left( \frac{i}{2\pi }\tilde{\theta }\theta  \right)Z\left( {\tilde{\alpha }} \right),\;\tilde{\alpha }\equiv \frac{1}{4{{\pi }^{2}}\alpha }
$},\nonumber \\
 \scalebox{0.85}{$\displaystyle\hspace{-30px}
   Z\left( \alpha  \right)\!\equiv\!\!\sum\limits_{n=-\infty }^{\infty }\!\!\exp\!!\left\{\frac{-\alpha }{2}{{\left( \theta\! -\!2\pi n \right)}^2}\!+\!in\tilde{\theta } \right\},\;Z\left( {\tilde{\alpha }} \right)\!\equiv\!\!\sum\limits_{\tilde{n}=-\infty }^{\infty }{\!\!\exp\!\left\{ \frac{-\tilde{\alpha }}{2}( \tilde{\theta }\!-\!2\pi \tilde{n} )^2\!-\!i\tilde{n}\theta\right\}}
 $},
\end{eqnarray}
Where $\tilde{\alpha }$ is a dimensionless energy dual to $\alpha$, and $Z\left(\alpha\right)$ and $Z\left(\tilde{\alpha } \right)$ are self dual partition functions to each other. The dual transformation in the Modified Villain formulation with two dimensions (two components) is as follows:
 \begin{eqnarray}\label{eqA12}
\scalebox{0.9}{$\displaystyle\hspace{-70px}
\sum\limits_{n_0,n_x=-\infty }^{\infty }\!\!\!\exp\!\left( -1 \right)\left\{ \frac{\alpha_0}{2}{\left(\nabla_{\tau }\theta \!-\!2\pi n_{\tau }\right)}^2\!+\!\frac{\alpha _x}{2}{\left( {{\nabla }_x}\theta\! -\!2\pi {n_x} \right)}^2\!+\!i\tilde{\theta }\omega  \right\}
$},\nonumber \\
 \scalebox{0.9}{$\displaystyle\hspace{-50px}
  \! =\!\frac{1}{2\pi \sqrt{{\alpha }_0{\alpha }_x}}\!\sum\limits_{\tilde{n}_{\tau },{\tilde{n}_x}=-\infty }^{\infty }\!\!\!\exp\!\left( -1 \right)\left\{ \frac{\tilde{\alpha }_0}{2}( {\nabla }_{\tau }\tilde{\theta }\!-\!2\pi\tilde{n}_{\tau })^2\!+\!\frac{\tilde{\alpha }_x}{2}( {\nabla }_x\tilde{\theta }\!-\!2\pi\tilde{n}_x)^2\!+\!i\theta \tilde{\omega } \right\}
 $},
\end{eqnarray}
where $\tilde{\omega }$ is defined as curl of $\tilde{n}_i$ \cite{ref11,ref14}, as follows:
 \begin{eqnarray}\label{eqA13}
\scalebox{1.0}{$\displaystyle
\hspace{100px} \tilde{\omega} \left( x,\tau  \right)\equiv{\varepsilon }_{ij}{\nabla }_i{\tilde{n}_j}\left( x,\tau  \right)
$},
\end{eqnarray}
where $\varepsilon_{\tau x}\!=\!-\varepsilon_{x\tau }\!=\!1$ is the Levi–Civita symbol of two dimensions. 
And $\tilde{\alpha }_{\tau }$ and $\tilde{\alpha }_x$ have the following crossed duality with respect to ${\alpha }_x$ and ${\alpha }_{\tau }$, respectively:
 \begin{eqnarray}\label{eqA14}
\scalebox{1.0}{$\displaystyle\hspace{50px}
\hspace{80px}\tilde{\alpha }_0\equiv \frac{1}{4{{\pi }^2}{\alpha_x}}
$},
\end{eqnarray}
 \begin{eqnarray}\label{eqA15}
\scalebox{1.0}{$\displaystyle\hspace{50px}
\hspace{80px}{\tilde{\alpha }_x}\equiv \frac{1}{4{{\pi }^2}{\alpha_0}}
$}.
\end{eqnarray}
Transforming the Villain approximation $Z_v$ in Eq. (\ref{eqA6}), adding the term $i\tilde{\theta }\omega$ for the dual phase $\tilde{\theta }$, and applying the Modified Villain formulation in two components in Eq. (\ref{eqA12}), we obtain
 \begin{eqnarray}\label{eqA16}
\scalebox{0.85}{$\displaystyle\hspace{-80px}
Z_{MV}\left( {\beta }_0,\beta  \right)\!\equiv\! R_{QV}\!\int\!\mathcal{D}\theta\int\!\mathcal{D}\tilde{\theta }\sum\limits_{\left\{ {n_i} \right\}}\exp\!\left( -1 \right)\!\sum\limits_{x,\tau }{\left\{ \frac{{\left({\beta }'_0\right)}_v}{2}{\biggl( {\nabla }_{\tau }\theta\! -\!2\pi n_{\tau } \biggr)}^2\!+\!\frac{{\left( {\beta }'_x \right)}_v}{2}{\biggl( {{\nabla }_x}\theta\! -\!2\pi n_x \biggr)}^2\!+i\tilde{\theta }\omega \right\}}
$},\nonumber \\
 \scalebox{0.85}{$\displaystyle\hspace{-70px}
  \!=\!C_{QV}\!\int\!\mathcal{D}\theta\!\int\mathcal{D}\tilde{\theta }\sum\limits_{\left\{\tilde{n}_i\right\}}\exp\! \left( -1 \right)\!\sum\limits_{x,\tau }{\left\{ \frac{{( \tilde{\beta }'_0 )}_v}{2}{( {{\nabla }_{\tau }}\tilde{\theta }\!-\!2\pi\tilde{n}_{\tau } )}^2\!+\!\frac{{( \tilde{\beta }'_x )}_v}{2}{( {\nabla }_x\tilde{\theta }\!-\!2\pi\tilde{n}_x )}^2\!+i\theta \tilde{\omega } \right\}}
 $},
\end{eqnarray}
where ${C_{QV}}\equiv {{\left[ {I_0}\left( {\beta }'_{0} \right){I_0}\left( {\beta }'_x \right) \right]}^{M_{\tau }M_x}}$ is normalization parameter, and Villain's inverse temperature ${( \tilde{\beta }_0 )}_v$ and ${( {\tilde{\beta }_x} )}_v$ are cross duals to ${\left( {\beta }_x \right)}_v$ and ${\left( {\beta }_0 \right)}_v$, respectively, as in Eq. (\ref{eqA14}), as follows: 
 \begin{eqnarray}\label{eqA17}
\scalebox{1.0}{$\displaystyle\hspace{100px}
( \tilde{\beta }'_0)_v\equiv \frac{1}{4{{\pi }^2}{\left( {\beta' }_x \right)}_v}
$},
\end{eqnarray}
 \begin{eqnarray}\label{eqA18}
\scalebox{1.0}{$\displaystyle\hspace{100px}
( \tilde{\beta }'_x )_v\equiv \frac{1}{4{\pi }^2{\left( {\beta' }_0 \right)}_v}
$}.
\end{eqnarray}
\section{Current of the momentum symmetry and the winding symmetry for $1+1d$ XY-model}
In this Appendix, we derive the Noether currents of momentum symmetry and winding symmetry according to references \cite{ref14} and \cite{ref15}.
From the partition function in Eq. (\ref{eqA15}), the dimensionless action $S'\left[ {\beta }_0, {\beta }_x \right]$, described in subspace rather than the entire $\tau$-x space, as follows: 
 \begin{eqnarray}\label{eqB1}
\scalebox{0.95}{$\displaystyle\hspace{-50px}
S'\left[ {\beta }_0,{{\beta }_x} \right]\!=\!\frac{{\left( {\beta }'_0\right)}_v}{2}\!\sum\limits_{ \tau-link}\!{\left( {\nabla }_{\tau }\theta\! -\!2\pi n_{\tau }\right)}^2\!+\!\frac{{{\left( {\beta }'_x \right)}_v}}{2}\!\sum\limits_{x-link}\!{\left( {\nabla }_x\theta \!-\!2\pi n_x\right)}^2\!+\!i\!\!\sum\limits_{plaquette}{\tilde{\theta }\omega }
$},
\end{eqnarray}
where $\sum\nolimits_{link}$ and $\sum\nolimits_{plaquette}$ are the sum of link and plaquette (or dual sites) in subspace of $1+2d$ space, respectively:
In the action of Eq. (\ref{eqB1}), there exists the following global symmetry for the phase field $\theta$ due to a real constant $c^m$ shift, called momentum symmetry: 
 \begin{eqnarray}\label{eqB2}
\scalebox{1.0}{$\displaystyle\hspace{100px}
\theta \to \theta +c^m
$}.
\end{eqnarray}
Such symmetries are called $U\left( 1 \right)$ momentum global symmetries according to string theory terminology.
From the Noether current of the momentum global symmetry in Eq. (\ref{eqB2}), the momentum current $J_i$\cite{ref13,ref14,ref20} is as follows: 
 \begin{eqnarray}\label{eqB3}
\scalebox{1.0}{$\displaystyle\hspace{50px}
J_i\left( x,\tau  \right)=-i{\left( {\beta }_i \right)}_v\left({\nabla }_i\theta -2\pi n_i \right)\left( x,\tau  \right)
$}.
\end{eqnarray}
Note that on the right-hand side of Eq. (\ref{eqB3}), no contraction is taken for the index $i$. In contrast, in the action of Eq. (\ref{eqB1}), there exists the following global symmetry for the real constant $c^w$ shift of the dual phase field $\tilde{\theta }$:
 \begin{eqnarray}\label{eqB4}
\scalebox{1.0}{$\displaystyle\hspace{100px}
\tilde{\theta }\to \tilde{\theta }+c^w
$}.
\end{eqnarray}
This symmetry is called global $U\left( 1 \right)$ winding symmetry. The winding current $J_i^w$ of symmetry in Eq. (B.4) is obtained as follows:
 \begin{eqnarray}\label{eqB5}
\scalebox{1.0}{$\displaystyle\hspace{50px}
J_i^w\left( x,\tau  \right)\equiv \frac{{\varepsilon }_{ij}}{2\pi }\Bigl[{\nabla }_{j}\theta \left( x,\tau  \right)-2\pi n_{j }\left( x,\tau  \right) \Bigr]
$},
\end{eqnarray}
From the partition function in Eq. (\ref{eqA15}), the dual dimensionless action $\tilde{S}'$ for the dimensionless action $S'$ in Eq. (\ref{eqB1}) is as follows:
 \begin{eqnarray}\label{eqB6}
\scalebox{0.9}{$\displaystyle\hspace{-60px}
\tilde{S}'[ \tilde{\beta }_0,\tilde\beta {x}]\!=\!\frac{( \tilde{\beta }'_0 )_v}{2}\!\!\!\!\sum\limits_{dual\;\tau-link}\!\!\left({\nabla }_{\tau }\tilde\theta\!-\!2\pi \tilde{n}_{\tau }\right)^2\!+\!\frac{( \tilde{\beta }'_x )_v}{2}\!\!\!\!\sum\limits_{dual\;x-link}\!\!\left({\nabla }_x\tilde\theta\!-\!2\pi \tilde{n}_x \right)^2\!+\!i\!\!\!\!\sum\limits_{dual\;plaquette}\!\!\!\!\theta\tilde\omega
$},
\end{eqnarray}
where $\sum\nolimits_{dual\;link}$ and $\sum\nolimits_{dual\;plaquette}$ are the sum of dual link and dual plaquette in $1+1d$ dual $\tau -x$ space, respectively.
From the Noether current for $U\left( 1 \right)$ momentum symmetry shifts $\tilde{\theta }\to \tilde{\theta }+c^m$ in dual $\tau -x$ space, the dual momentum current ${\tilde{J}}_{\mu }$\cite{ref13,ref14,ref20} is as follows:
 \begin{eqnarray}\label{eqB7}
\scalebox{1.0}{$\displaystyle\hspace{40px}
\tilde{J}_{i}\left( x,\tau  \right)=-i(\tilde{\beta}_{i})_v\left( {{\nabla }_{i}}\tilde{\theta }-2\pi\tilde{n}_{i} \right)\left( x,\tau  \right)
$},
\end{eqnarray}
Similarly, from the Noether current for the shift $\theta \to \theta +c^w$ of global $U\left( 1 \right)$ winding symmetry in the dual $\tau -x$, we define the dual winding current $\tilde{J}_{\mu }^w$ as follows:
 \begin{eqnarray}\label{eqB8}
\scalebox{1.0}{$\displaystyle\hspace{50px}
\tilde{J}_{i}^w\left( x,\tau  \right)\equiv \frac{{\varepsilon }_{ij }}{2\pi }\left[ {\nabla }_{j}\tilde{\theta }\left( x,\tau  \right)-2\pi \tilde{n}_{j}\left( x,\tau  \right) \right]
$}.
\end{eqnarray}
\section{Algorithm for the DHM}
In this appendix, we outline the DHM, an algorithm for constructing self-dual systems developed in our previous paper \cite{ref22,ref23}. First, put the single JJ Hamiltonian $H$ as follows: 
 \begin{eqnarray}\label{eqC1}
\scalebox{1.0}{$\displaystyle\hspace{40px}
H\left( \theta ,N \right)=E_cN^2+E_J\left( 1-\cos \Delta \theta  \right)
$},
\end{eqnarray}
where, $E_c\equiv \left( 2e \right)^2/{2C}\;$ is charging energy per Cooper pair, therefore, $E_J\equiv {\Phi }_0{I_c}/{2\pi }\;$ is the Josephson energy, $I_c$ and ${\Phi }_0=h/2e\;$ are the critical current and the magnetic flux- quantum, respectively, and $N$ and $\Delta \theta$ are the particle number of the Cooper pair and the phase difference of the Cooper pair, respectively. Since $\hbar N$ and $\Delta \theta$ are canonically conjugate to each other, Poisson bracket $\left\{ \hbar N,\Delta \theta  \right\}_p\!=\!1\;$is satisfied. From the equation of motion for the single JJ Hamiltonian in Eq. (\ref{eqC1}), the following Josephson equation can be derived.
 \begin{eqnarray}\label{eqC2}
\scalebox{1.0}{$\displaystyle\hspace{20px}
V\left( t \right)=\frac{\hbar }{2e}\frac{\partial \Delta \theta }{\partial t}=\frac{2N}{2e}{E_c},\;I\left( t \right)=2e\frac{\partial N}{\partial t}=-\frac{2\pi }{{\Phi }_0}E_J\sin \Delta \theta
$},
\end{eqnarray}
where $V\left( t \right)$ and $I\left( t \right)$ are the voltage and the current for the JJ, respectively. Next, put a single QPSJ \cite{ref22,ref23,ref24,ref25,ref26,ref27} Hamiltonian $\tilde{H}$ as follow:
 \begin{eqnarray}\label{eqC3}
\scalebox{1.0}{$\displaystyle\hspace{60px}
\tilde{H}( \tilde{\theta },\tilde{N} )={E_L}{\tilde{N}}^2+{E_S}( 1-\cos \Delta \tilde{\theta } )
$},
\end{eqnarray}
where, $E_L\!\equiv\!{\Phi_0}^2/2L$ is the inductive energy per magnetic flux quantum, $E_S\!\equiv\!2eV_c/2\pi\;$is the QPS amplitude, $V_c$ is the critical voltage, ${Z}'_{DAXY}$ and $\Delta \tilde{\theta }$ are the particle number of magnetic flux-quantum and the phase difference of magnetic flux-quantum in QPS junction respectively.
Since $\hbar \tilde{N}$ and $\Delta\tilde{\theta }$ are canonically conjugate to each other, Poisson bracket $\{ \hbar \tilde{N},\Delta \tilde\theta\}_p=1$ is satisfied. From the equation of motion for the single QPSJ Hamiltonian in Eq. (\ref{eqC3}), the following dual Josephson equation \cite{ref22,ref23} can be derived. 
 \begin{eqnarray}\label{eqC4}
\scalebox{1.0}{$\displaystyle\hspace{20px}
\tilde{V}\left( t \right)=\frac{\hbar }{\Phi_0}\frac{\partial \Delta \tilde{\theta }}{\partial t}=\frac{2\tilde{N}}{\Phi_0}E_L,\;\tilde{I}\left( t \right)=-\Phi_0\frac{\partial \tilde{N}}{\partial t}=\frac{2\pi }{2e}E_S\sin \Delta \tilde{\theta }
$},
\end{eqnarray}
where $\tilde{V}\left( t \right)$ and $\tilde{I}\left( t \right)$ are the dual voltage and the dual current for the dual JJ \cite{ref22}, respectively. 
As a first step of the DH method, we assume the following two conditions between the Josephson equation in Eq. (\ref{eqC2}) and the dual Josephson equation in Eq. (\ref{eqC4}):
\begin{eqnarray}\label{eqC5}
\scalebox{1.0}{$\displaystyle\hspace{100px}
V\left( t \right)\equiv \tilde{I}\left( t \right),\;I\left( t \right)=\tilde{V}\left( t \right)
$},
\end{eqnarray}
The two conditions of Eq. (\ref{eqC5}) are called self dual conditions. The self dual conditions \cite{ref22,ref23} in Eq. (\ref{eqC5}) is known as the duality principle of electric circuits, and it is known that similar operations can be satisfied by swapping the roles of resistance and conductance, inductance and capacitance, and current and voltage in a classical electric circuit. The next step in the DHM is to derive a relational expression between canonical conjugate variables that are dual to each other according to the self dual conditions of Eq. (\ref{eqC5}). Imposing the conditions of Eq. (\ref{eqC5}) between Eq. (\ref{eqC2}) and (\ref{eqC4}), we obtain the following two relations \cite{ref22} between the phase difference and the number of particles between systems that are dual to each other.
\begin{eqnarray}\label{eqC6}
\scalebox{1.0}{$\displaystyle\hspace{50px}
\tilde{N}\left( t \right)=\frac{-1}{2\pi }\sin \Delta \theta \left( t \right),\;N\left( t \right)=\frac{1}{2\pi }\sin \Delta \tilde{\theta }\left( t \right)
$},
\end{eqnarray}
The linear approximation of Eq. (\ref{eqC6}) is a well-known relationship between the phase and the number of particles, as shown in the following equations: 
\begin{eqnarray}\label{eqC7}
\scalebox{1.0}{$\displaystyle\hspace{40px}
\tilde{N}\left( t \right)=\frac{\Phi }{{\Phi }_0}=\frac{-1}{2\pi }\Delta \theta \left( t \right),\;N\left( t \right)=\frac{Q}{2e}=\frac{1}{2\pi }\Delta \tilde{\theta }\left( t \right)
$},
\end{eqnarray}
The right-hand sides of Eq. (\ref{eqC6}) and (\ref{eqC7}) correspond to the winding currents in Eq. (\ref{eqB7}) and (\ref{eqB10}) introduced in Appendix B. 
If it is recognized that the relationships described in Eq. (\ref{eqC6}) are satisfied, the relationship between the $QPS$ amplitude and charging energy per single-charge, and the relationship between Josephson energy and inductive energy per magnetic flux-quantum, are as follows \cite{ref22}:
\begin{eqnarray}\label{eqC8}
\scalebox{1.0}{$\displaystyle\hspace{70px}
E_S=\frac{1}{2{\pi }^2}{E_c},\;E_J=\frac{1}{2{\pi }^2}{E_L}
$},
\end{eqnarray}
Furthermore, inductance and capacitance are related to the critical current $I_c$ and the critical voltage $V_c$, respectively, as follows \cite{ref22}:
\begin{eqnarray}\label{eqC9}
\scalebox{1.0}{$\displaystyle\hspace{80px}
L=\frac{{\Phi }_0}{2\pi I_c},\;C=\frac{2e}{2\pi V_c}
$},
\end{eqnarray}
By the Josephson current $I$ and the dual Josephson current $\tilde{I}$ in Eq. (\ref{eqC2}) and (\ref{eqC4}), the particle number $N$ of the cooper pair and the particle number $\tilde{N}$ of the magnetic flux-quantum can be expressed as follows, respectively:
\begin{eqnarray}\label{eqC10}
\scalebox{1.0}{$\displaystyle\hspace{60px}
\tilde{N}\left( t \right)=\frac{{\Phi }_0}{{\left( 2\pi  \right)}^2E_J}I\left( t \right),\;N\left( t \right)=\frac{2e}{{\left( 2\pi  \right)}^2E_S}\tilde{I}\left( t \right)
$},
\end{eqnarray}
To summarize the results of this appendix, by accepting the results of Eq. (\ref{eqC6}) to (\ref{eqC9}) obtained under the double condition of Eq. (\ref{eqC5}), the Hamiltonian in Eq. (\ref{eqC1}) and (\ref{eqC3}), it was found that its duality was completely guaranteed. Among the results, Eq. (\ref{eqC6}) is particularly important as it becomes the starting point as a relational expression for creating a self dual system in the section 3. Eq. (\ref{eqC10}) is also very important for constructing the dual transformation of a self-dual system because it yields the relationship between the number of particles and the current between systems that are dual to each other.
\section{Dual transformation by the SVF for 1+1d XY-model}
In this Appendix. we follow the method of our previous paper \cite{ref22}, we show that the result of the dual transformation by the SVF in the $1+1d$XY-model is self dual as well as the result by the two-component MVF. The starting point of the theory is the partition function of the SVF \cite{ref7,ref8,ref9,ref10} in Eq. (\ref{eqA6}). For Eq. (\ref{eqA6}), introduce the Poisson's summation formula ( modular identity of the Jacobi theta function ) \cite{ref9,ref10} as follows:
\begin{eqnarray}\label{eqD1}
\scalebox{1.0}{$\displaystyle\hspace{50px}
\sum\limits_{n=-\infty }^{\infty }e^{\frac{-\alpha }{2}{\left( \theta -2\pi n \right)}^2}=\frac{1}{\sqrt{2\pi \alpha }}\sum\limits_{b=-\infty }^{\infty }e^{\frac{-1}{2\alpha }b^2+ib\theta }
$},
\end{eqnarray}
As a result, the partition function in Eq. (\ref{eqA6}) can be rewritten as follows \cite{ref10,ref22}: 
\begin{eqnarray}\label{eqD2}
\scalebox{0.9}{$\displaystyle\hspace{-10px}
Z_v\left(\beta_0,\beta_x \right)\equiv C_v\sum\limits_{\left\{ b \right\}}\delta_{\nabla_j b_j,0}\exp\sum\limits_{x,\tau }\left\{ \frac{-b_0^2\left( x,\tau  \right)}{2\left( \beta'_0\right)_v}+\frac{-b_x^2\left( x,\tau  \right)}{2\left( {\beta'_x} \right)_v} \right\}
$},
\end{eqnarray}
where $C_v\!\equiv\!\left[ I_0\left(\beta'_0\right)I_0\left(\beta'_x \right) \right]^{M_{\tau }M_x}$ is normalization parameter, and $b_i\!\left( x,\tau  \right)$$\left( i=\tau ,x \right)$ are integer value auxiliary magnetic fields where the condition ${\nabla }\!_{\tau }b_{\tau}+{\nabla }\!_{x }b_x\!=\!0$ is satisfied. The partition function in Eq. (\ref{eqD2}) and (\ref{eqA6}) is dual relation, but not yet self dual relation. Dual integer value magnetic field $\tilde{b}_i\left( x,\tau  \right)$ is introduced to integer value auxiliary magnetic field $b_i\left( x,\tau  \right)$, as follows \cite{ref22}: 
\begin{eqnarray}\label{eqD3}
\scalebox{1.0}{$\displaystyle\hspace{100px}
b_i\left( x,\tau  \right)\equiv{\varepsilon }_{ij}{\tilde{b}}_j\left( x,\tau  \right)
$},
\end{eqnarray}
where ${\varepsilon }_{0x}=-{\varepsilon }_{x0}=1$ is the Levi-Civivita symbol of two dimensions. By using the dual transformations of Eq. (\ref{eqD3}), the following equations are obtained for Eq. (\ref{eqD2}): 
\begin{eqnarray}\label{eqD4}
\scalebox{0.9}{$\displaystyle\hspace{-10px}
 Z_v\left( {\beta }_0,{\beta }_x \right)\equiv C_v\sum\limits_{\left\{ \tilde{b} \right\}}{{{\delta }_{{{\varepsilon }_{ij}{\nabla }_i\tilde{b}_j},0}}\exp }\sum\limits_{x,\tau }{\left\{ \frac{-{\tilde{b}_0}^2\left( x,\tau  \right)}{2{\left( {\beta }'_x \right)}_v}+\frac{-{\tilde{b}_x}^2\left( x,\tau  \right)}{2{{\left( {\beta }'_0 \right)}_v}} \right\}}
$},
\end{eqnarray}
Introducing Poisson's formula in Eq.(\ref{eqD5}) below into Eq.(\ref{eqD4}) and rewriting the integer value field $\tilde{b}_i\left( x,\tau  \right)$ into a continuous value field $\tilde{B}_i\left( x,\tau  \right)$, we obtain the following Eq.(\ref{eqD6}) \cite{ref22}. 
\begin{eqnarray}\label{eqD5}
\scalebox{0.88}{$\displaystyle\hspace{-20px}
 \sum\limits_{\left\{ \tilde{b}\right\}}{{{\delta }_{{\varepsilon }_{ij}{\nabla }_i\tilde{b}_j,0}}\left( \cdot \cdot \cdot  \right)}=\int\mathcal{D}{\tilde{B}}_j\left( \cdot \cdot \cdot  \right)\sum\limits_{\left\{ l \right\}}{{\delta }_{{\varepsilon }_{ij}{\nabla }_i{l_j},0}}\exp \sum{\left( i2\pi{\varepsilon }_{ij}{\tilde{B}_i}l_j \right)}
$},
\end{eqnarray}
\begin{eqnarray}\label{eqD6}
\scalebox{0.88}{$\displaystyle\hspace{-60px}
Z_v\left(\beta_0,\beta_x \right)=C_v\sum\limits_{\left\{l\right\}}\delta_{\varepsilon_{ij}\nabla_il_j,0}\int \mathcal{D}\tilde{B}_j\exp \sum\limits_{x,\tau }{\left\{ \frac{-{\tilde{B}_0}^2}{2\left( \beta'_x \right)_v}+\frac{-{\tilde{B}_x}^2}{2\left( \beta'_0\right)_v}+i2\pi l_x\tilde{B}_0-i2\pi l_0\tilde{B}_x \right\}}
$},
\end{eqnarray}
where $l_i\left( x,\tau  \right)$ corresponds to integer value current. Integrating over the continuous value field $\tilde{B}_i$ of Eq. (\ref{eqD6}) yield the following equation:
\begin{eqnarray}\label{eqD7}
\scalebox{0.9}{$\displaystyle\hspace{20px}
Z_v=R_{QV}\sum\limits_{\left\{l\right\}}\delta_{\varepsilon_{ij}\nabla_il_j,0}\exp \left( -1 \right)\sum\limits_{x,\tau }{\left\{ 2{\pi}^2\left(\beta'_0 \right)_v{l_0}^2+2{\pi }^2\left(\beta'_x \right)_v{l_x}^2 \right\}}   
$},
\end{eqnarray}
The Kronecker delta, when rewritten in the integral form, allow the equations to be written as follows: 
\begin{eqnarray}\label{eqD8}
\scalebox{0.9}{$\displaystyle\hspace{0px}
Z_v=R_{QV}\sum\limits_{\left\{ l \right\}}{\int\mathcal{D}\tilde{\theta }}\exp \left( -1 \right)\sum\limits_{x,\tau }{\left\{ 2{\pi }^2{\left( {\beta }'_0 \right)}_v{l_0}^2+2{{\pi }^2}{\left( {\beta }'_x \right)}_v{l_x}^2+i\tilde{\theta }{\omega }' \right\}} 
$},
\end{eqnarray}
where , the third term $i\tilde{\theta }{\omega }'$ of the dimensionless action in Eq. (\ref{eqD8}) can be considered to correspond to the additional term  in Eq. (A.8) in the MVF. $\omega$ is defined as the curl of $l_i$, as follows \cite{ref22}: 
\begin{eqnarray}\label{eqD9}
\scalebox{1.0}{$\displaystyle\hspace{100px}
{\omega }'\left( x,\tau  \right)\equiv {\varepsilon }_{ij}{{\nabla }_i}l_j\left( x,\tau  \right)
$},
\end{eqnarray}
Using the identity of Eq. (\ref{eqD1}) for Eq. (\ref{eqD8}), the equations become \cite{ref22}:
\begin{eqnarray}\label{eqD10}
\scalebox{0.9}{$\displaystyle\hspace{-10px}
Z_v=C_v\sum\limits_{\left\{ \tilde{n} \right\}}{\int{D\tilde{\theta }}}\exp \sum\limits_{x,\tau }{\left\{ \frac{-{{( {\nabla }_{\tau }\tilde{\theta }-2\pi \tilde{n}_0 )}^2}}{8{{\pi }^2}{{\left( {\beta }'_x \right)}_v}}-\frac{( {\nabla }_x\tilde{\theta }-2\pi \tilde{n}_x )^2}{8{{\pi }^2}{\left( {\beta }'_0 \right)}_v} \right\}}
$},
\end{eqnarray}
Imposing the same relation $( \tilde{\beta }'_0)_v=1/4{\pi }^2( {\beta }'_x )_v$ and $( \tilde{\beta }'_x)_v=1/4{\pi }^2\left( {\beta }'_0 \right)_v$ as in Eq. (\ref{eqA17}) and (\ref{eqA18}) on Eq. (\ref{eqD10}), the self dual relation is established as in Eq. (\ref{eqA16}), as follows \cite{ref22}:
 \begin{eqnarray}\label{eqD11}
\scalebox{0.86}{$\displaystyle\hspace{-80px}
Z_V\left( {\beta }_0,{\beta }_x \right)={R}_{QV}\int{D\theta }\sum\limits_{\left\{ n \right\}}{\exp }\sum\limits_{x,\tau }{\left\{ \frac{-\left( {\beta }'_0 \right)_v}{2}\left( {{\nabla }_{\tau }}\theta -2\pi {n_{\tau }} \right)^2-\frac{\left( {\beta }'_x \right)_v}{2}\left( {\nabla }_x\theta -2\pi n_x \right)^2 \right\}}
$},\nonumber \\
 \scalebox{0.86}{$\displaystyle\hspace{-80px}
  \!=\!C_v\sum\limits_{\left\{ \tilde{n}\right\}}\int{D\tilde{\theta }}\exp \sum\limits_{x,\tau }{\left\{ \frac{-{( \tilde{\beta' }_0 )}_v}{2}( {\nabla }_{\tau}\tilde{\theta }\!-\!2\pi {\tilde{n}_0} )^2\!-\!\frac{(\tilde{\beta' }_x )_v}{2}( {\nabla }_x\tilde{\theta }\!-\!2\pi \tilde{n}_x)^2\right\}}\!\equiv\! {C_v}{\tilde{R}}^{-1}_{QV}\tilde{Z}_V( \tilde{\beta }_0,\tilde{\beta}_x)
 $},
\end{eqnarray}
\begin{eqnarray}\label{eqD12}
\scalebox{0.9}{$\displaystyle\hspace{-80px}
{\tilde{Z}}_V(\tilde{\beta}_0,\tilde{\beta}_x)\equiv{\tilde{R}_{QV}}\sum\limits_{\left\{\tilde{n}\right\}}{\int{D\tilde{\theta}}}\exp\sum\limits_{x,\tau}{\left\{\frac{-{({\tilde{\beta}'_0})}_v}{2}{{\left({{\nabla}_{\tau}}\tilde{\theta}-2\pi{\tilde{n}_0}\right)}^2}-\frac{{({\tilde{\beta}'_x})}_v}{2}{{\left({{\nabla}_x}\tilde{\theta}-2\pi{\tilde{n}_x}\right)}^2}\right\}}
$},\nonumber \\
 \scalebox{0.95}{$\displaystyle\hspace{80px}
 \tilde{R}_{QV}\equiv[R_v(\tilde{\beta' }_0 )R_v(\tilde{\beta' }_x ) ]^{M_{\tau }M_x}
 $}.
\end{eqnarray}

\def\thesection{}


\end{document}